\documentclass[12pt,preprint]{aastex}

\usepackage{natbib}
\usepackage{mathtools}
\pdfoutput=1

\newcommand{\mybf}{}

\usepackage{graphicx}

\begin{document}
\author{A. Dorodnitsyn\altaffilmark{1,2}, T. Kallman\altaffilmark{1}
}
\altaffiltext{1}{Laboratory for High Energy Astrophysics, NASA Goddard Space Flight Center, Code 662, Greenbelt, MD, 20771, USA}
\altaffiltext{2}{Department of Astronomy/CRESST, University of Maryland, College Park, MD 20742, USA}

\title{
AGN obscuration from winds: from dusty infrared-driven to warm and X-ray photoionized}
\begin{abstract}

\end{abstract}
We present calculations of AGN winds at $\sim$parsec scales, along with the associated obscuration.
We take into account the pressure of infrared radiation on dust grains and the interaction of X-rays from a central black hole with hot and cold plasma. Infrared radiation (IR) is incorporated in radiation-hydrodynamic simulations adopting the flux-limited diffusion approximation.
We find that in the range of X-ray luminosities L=0.05 - 0.6$L_{\rm edd}$,
the Compton-thick part of the flow (aka torus) has an opening angle of approximately $72^{\circ}-75^{\circ}$ regardless of the luminosity. At $L\gtrsim 0.1\,L_{\rm edd}$ the outflowing dusty wind provides the obscuration
with IR pressure playing a major role.
The global flow consists of two phases:
the cold flow at inclinations $\theta\gtrsim 70^{\circ}$ and a hot, ionized wind of lower density at lower inclinations. The dynamical pressure of the hot wind is important in shaping the denser IR supported flow.
At luminosities $\leq$0.1$L_{\rm edd}$ episodes of outflow are followed by extended periods when the wind switches to slow accretion.

\section{Introduction}
Active galactic nuclei (AGNs) are among the most fascinating objects in the universe:
various energetic processes interplay in a volume a few cubic parsecs around a supermassive black hole (BH); radiation is likely decisive in connecting the accretion and feedback modes of AGN.
Dusty, radiation driven flows are very effective 
in removing large masses of accreting gas, returning  
it back the galaxy, and naturally limiting the growth rate of a supermassive BH by the feedback with the host galaxy. 
Radiation also provides a non-linear causal connection between distant parts of an AGN itself, coupling together vastly different physical scales. 
Despite the pivotal role of radiation in the first principle modeling of AGN and the significant progress in theoretical modeling over the last 10 years, proper incorporation of radiation into 
numerical models still presents a significant computational challenge. 

Well established phenomenology places AGNs into two distinct classes which roughly differ by the presence (type I) or absence (type II) of the broad emission lines in the optical and UV spectra. 
According to AGN unification paradigm \citep{Rowan-Robinson77, AntonucciMiller1985, UrryPadovani95}, this dichotomy is an artifact of a pure geometrical origin. 
In order to be observed as type II object, an AGN must be viewed at high inclination, in extreme cases close to the equatorial plane, while in the case of a type I AGN the observer's line of sight 
is at lower inclination, i.e. closer to the symmetry axis.
The presence of geometrically thick toroidal obscuration is pivotal to the unification scheme of AGNs.

The discovery of broad permitted emission lines in the polarized spectrum of the nearby Seyfert II galaxy NGC 1068 
\cite{Antonucci84}, and \cite{AntonucciMiller1985}, provides evidence that a type I nucleus is harbored inside the obscuring ring of matter.
The observed polarization is likely due to scattering of light emitted by the type-I core 
and then being scattered within an outflow emanating from the same nucleus.

Observations of AGNs and quasars reveal that dust in large quantities can be present within a few parsecs from the BH, and a significant part of X-rays and UV are reprocessed into infrared (IR).
For example, in luminous quasars almost half of the bolometric luminosity can be reprocessed into IR. This results in
a 1-100$\mu{\rm m}$ ''hump'' in the quasar spectral energy distribution (SED) \cite{Richards06}. 
In nearby AGNs
multi-temperature dust distributions are observed directly by the Very Large Telescope Interferometer (VLTI). Such mid-infrared observations of the prototypical Seyfert II galaxy NGC 1068 \citep{Jaffe2004} and the Circinus galaxy \citep{Tristram07} depicts a picture of a dusty, clumpy plasma exposed to external illumination.

{\mybf
Various models have been proposed to explain the real or apparent thickness of the torus:

1) a ''warped disk model'', these are global warps in a locally geometrically thin accretion disk
{\citep[i.e.,][]{Phinney89, Sanders89}, 

2) i) models of a quasi-static, rotating torus involving ''clouds'' \citep[i.e.,][]{KrolikBegelman88,BeckertDuschl04,Nenkova08}.
Vertical thickness of the torus is supported via large velocity dispersion of clouds, and small-scale magnetic field is required to maintain elasticity of clouds during collisions; ii) models which rely on infrared radiation pressure on dust to keep rotating torus vertically thick \citep{Krolik07}. Most of IR radiation in this models comes from reprocessing of external X-ray, and UV radiation.

3) a ''wind torus''
A collar of outflowing matter may provide obscuration.
i) MHD wind: possibility is that a global magnetic field can be involved in the form of an MHD wind \citep{KoniglKartje94,ElitzurShlosman06} or directly supporting a quasi-static torus \citep{Lovelace98}, or in a combination of radiative and magnetic driving \citep{Emmering92,Everett05, Keating12}; note that sufficiently strong (of the order of the equipartition one) poloidal magnetic field can also help to solve the problem of how the angular momentum is removed from accreting matter, however the origin of such highly ordered field remains unclear
ii) an outflow (or failed wind) driven by IR radiation
pressure on dust \citep{Dorodnitsyn11a, Dorodnitsyn12a}.
All these models have their benefits and drawbacks.
}
Models investigating the properties of a self-gravitating, non-radiative, dissipationless torus through N-body simulations have been made by \cite{Bannikova12}.

The accreting dusty gas at $\sim$parsec distances from the BH is susceptible to self-gravitating instability; since it is exposed to intense radiation strong gravito-thermal instabilities can develop.  

In the outflowing wind, the non-linear phase of thermal instability 
is associated with ''clumps'' \citep[i.e.,][]{Elitzur08}. Presumably such clouds are confined by small scale magnetic field.
Whether they are secondary to the flow and behave more like ''waves'', i.e. temporary structures evolving on a dynamical or radiation times scales, or real long-lasting ''clumps'' remains unresolved.
All models involving clouds need to explain how to avoid too much dissipation through the cloud-cloud collisions, which would otherwise raise the temperature to a fraction of the virial temperature $T_{\rm vir,g}$. In the present studies we do not consider the effect of magnetic field (both ordered and chaotic).

It has long been recognized that the geometrical thickness of the obscuration presents a significant problem. 
In order to explain why the torus is geometrically thick, models have been constructed which rely on gas pressure, magnetic forces, and on radiation pressure either alone or in combination with other mechanisms. 
If one relies on gas pressure to support vertical thickness of the torus then virial arguments suggest that the temperature of such gas should be of the order of 
$T_{\rm vir,g}=2.6 \times 10^{6}\,M_{7}/r_{\rm pc}$K, where $M_{7}$ is the black hole (BH) mass in $10^{7}\,M_{\rm \odot}$, and $r_{\rm pc}$ is the distance in parsecs. Such high temperatures are not compatible with the existence of dust.

As mentioned before,
the presence of a small scale magnetic field can provide necessary elasticity to cloud-cloud collisions. 
However, this argument is not without a flaw. It is natural to assume that turbulence develops. 
If so,  this leads to the magnetic field being of the order of the equipartition field.
A result is that the dissipation of magneto-sonic waves will raise the electron temperature to be of the order of $T_{\rm vir,g}$.

\mybf{
The connection between a toroidal obscuration, UV absorption and warm absorbers can be self-consistently assessed by global multidimensional simulations. The prequisite for such an endeavor is to include various physical effects and couple them to radiation. Additionally why such calculations are currently beyond the reach is that high numerical resolution is needed to resolve multiphase behavior of the outflowing gas. These are the reasons why different works including present studies
assess smaller aspects of the bigger picture. 

The connection between a Broad Line Region and dusty outflow was considered by
\cite{CzernyHryniewicz11}. They suggested that the gas of the dusty accretion disk is weekly coupled to the BH due to radiation pressure on dust resulting in a
dusty wind which gives the low-ionization part of the broad line region. 
\cite{Bottorff2000} considered  the relation between the observed properties of UV and X-ray absorbers with those predicted from a magnetohydrodynamical self-similar wind model. They conclude that in case of the 
well-studied Seyfert 1 galaxy NGC 5548 the intrinsically clumped absorber is biased toward smaller radii in case of warm absorber and are likely located at higher distances in case of UV absorption features.
} }
 
Other models are based on the possibility that the pressure of the infrared radiation on dust may itself be enough to support the vertical thickness of the torus. 
The idea of a static infrared support was explored in \cite{Krolik07} where an approximate analytical model for the torus was constructed showing that there is enough radiation force to support a static 
vertically thick torus. However, it was argued in \cite{Dorodnitsyn11a} (Paper I), that it is very unlikely that the IR supported torus would be static. It was shown that the equilibrium between rotational, gravitational and radiation forces cannot be maintained
if the temperature in the torus exceeds that of  
$T_{\rm vir, r}   \simeq 312\,(n/10^{5} M_7 {r_{\rm pc}^{-1} })^{1/4} - 987 \,(n/10^{7} M_7 {r_{\rm pc}^{-1} })^{1/4} \,{\rm K}$, where $n$ is the number density,  necessarily resulting in the torus being an outflow.
In the following paper {\citep[][Paper II]{Dorodnitsyn12a} we supported this idea with 
radiation-hydrodynamics simulations of the outflowing accretion disk wind, driven by the radiation pressure on dust. The distribution of IR radiation was calculated using a flux-limited diffusion approximation.

An important simplification was made in Papers I and II: Only the infrared-driven part of the wind was considered.
The argument was that this portion of the wind is far enough from the BH so all the conversion of X-rays to IR already happened outside the computational domain.
That allowed us to consider only one group of radiation (the IR). The radiation temperature at the boundary was calculated to match the input energetics of soft X-rays. In other words we did not treat explicitly the reprocessing of soft X-rays into IR. Correspondingly, we were not able to follow the interaction of the  X-ray heated hot component of the flow with the cold, dusty IR supported component.

In this paper we calculate the wind structure and explicitly treat the interaction between X-rays and gas with subsequent conversion of X-rays to IR. This allows us to follow in dynamics the transition from hot, photoionized to cold, dusty and IR supported flow. To make calculations numerically tractable we made several major simplifications in treatment of the transfer of X-rays 
as well as their interaction with the gas. For example, 
transfer of X-rays is reduced to simple attenuation with no multiple scatterings taken into account.
and we approximately calculate the heating of dust grains by X-rays. The rest of our computational radiation-hydrodynamics framework remains unchanged from Paper II:
the transfer of the infrared radiation and its interaction with matter is treated in a flux-limited diffusion approximation. The calculations are time-dependent and 2.5D, i.e.  3  dimensions in cylindrical coordinates  and assuming axial symmetry.

The plan of this paper is as follows:
After this introduction
the flow of this paper goes through Section 2 where we review 
basic assumptions of our model, including the description of the radiating fluid of the 
AGN dusty wind by means of radiation hydrodynamics equations. 
Interaction of the cold and hot components of the wind with
X-rays are described in Section 3 including corresponding approximations. In Section
4 we describe the numerical setup, initial and boundary conditions. The results are summarized in Section 5 while the final discussion is postponed to Conclusions, where we
highlight prospects and limitations of our approach to the problem of AGN winds.

\section{Assumptions and physical model}

The flow in which the continuum radiation plays an important role is described by the system of radiation hydrodynamics equations.
To first order in $v/c$, these can be formulated in the following form
\citep{MihalasBookRadHydro}:

\begin{eqnarray}
D_{t}\rho &+& \rho\,{\bf \nabla\cdot v}=0\mbox{,}\label{eq11}\\
D_{t}\bf {v} &=& -\frac{1}{\rho}\,\nabla p + {\bf g_{\rm rad}} -\nabla \Phi \mbox{,}\label{eq12}\\
\rho D_{t}\left(\frac{e}{\rho}\right) & = & - p{\bf \nabla\cdot v} - 4\pi\chi_{\rm P}B + c\chi_{\rm E}E
+n^{2} (\Gamma-\Lambda)
\mbox{,}\label{eq13}\\
\rho D_{t}\left(\frac{E}{\rho}\right) & = & -{\bf \nabla\cdot F} - {\bf \nabla v : P} + 4\pi\chi_{\rm P}B- c\chi_{\rm E}E\mbox{,}\label{eq14}
\end{eqnarray}
where quantities related to matter: $\rho$, $p$, and $e$ are
the material mass density, gas pressure and gas energy density, ${\bf v}$ is the velocity;
quantities related to radiation are the frequency-integrated moments (for definitions see Appendix A):  $E$ is the radiation energy density, 
$\bf F$ is the radiation flux, and $\bf P$ is the radiation pressure tensor; $\chi_{\rm P}$, $\chi_{\rm E}$ are the Planck
mean and energy mean absorption opacities (in $\rm cm^{-1}$),
$c$ is the speed of light, $B=\sigma T^{4}/\pi$ is the Planck function, and 
$\sigma = a c/4$ is the Stefan-Boltzmann constant, $T$ is the gas temperature;
other notation include the convective derivative,
$ D_{t} = \frac{\partial}{\partial t} + \bf{v\cdot \nabla}$, and ${\bf \nabla v : P}$ denotes
the contraction $(\partial_{j} v_{i}) P^{ij}$. Notice that all the dependent variables in 
(\ref{eq11})-(\ref{eq14}) are evaluated in the co-moving frame. Notice that $E=aT_{\rm r}^{4}$, where $T_{\rm r}$ is the radiation temperature. Interaction between X-rays and the gas is incorporated in the last term of the gas energy equation
(\ref{eq13}), where $\Gamma$ and $\Lambda$ are respectively heating and cooling rates.
The equation of state for the gas is assumed to be polytropic: $p=K\rho\gamma$, where 
$\gamma = 1 + 1/n$, and $n$ is the polytrope index, and $p = (\gamma-1) e$. 
The gas is one-component, one-temperature 
$T = p \mu /\rho {\cal R}$, where $\mu$ is the mean molecular weight per particle, 
${\cal R} = 8.31\times 10^{7} {\rm erg\, K^{-1} \, g^{-1}}$ is the
universal gas constant and plasma with $\gamma = 5/3$ is assumed to constitute the flow. 
Three components of the flow velocity $v_{R}$, $v_{z}$, and $v_{\phi}$ are calculated in cylindrical
coordinates $z$, $R$, assuming azimuthal symmetry, $\partial _\phi = 0 $.

Frequency-independent moments 
$E$, $\bf F$,  which appear in the above set of RHD equations are obtained by
calculating angular moments from the frequency-integrated specific intensity, $I({\bf r}, \Omega, \nu, t)$ and are given in Appendix A. 
Forces which are explicitly taken into account include radiation, ${\bf g_{\rm rad}}$ pressure and gas pressure, $\rho^{-1}\nabla p$.

These equations should be supplemented by
the radiation force, ${\bf g_{\rm rad}}$, which is calculated from the following relation	

\begin{equation}\label{radforce}
{\bf g_{\rm rad}}	= \frac{1}{c}\frac{\bf \chi_{\rm F} \,{\bf F} }{\rho}\mbox{,}
\end{equation}
where $\chi_{\rm F}= \chi_{\rm a}+\chi_{\rm T}$ is the total flux mean opacity consisting of absorption opacity, $\chi_{\rm a}$ and the Thomson scattering opacity, $\chi_{\rm T}$. 
As in Papers I, and II we did not differentiate between $\chi_{\rm F}$, $\chi_{\rm P}$ and $\chi_{\rm E}$.
Radiation pressure on dust is the most important from the dynamical point of view (Paper I).
Notice that in the infrared,
the Rosseland mean opacity, $\kappa_{\rm d}=\chi_{\rm d}/\rho$ of dust with 
the temperature $10^{2} -10^{3}$ K,
is approximately $10-30$ times larger than that of the electron Thomson opacities \citep{Semenov03}.  
In this paper when calculating radiation pressure force in (\ref{radforce}) we take into account only IR pressure on dust.
Including UV force (in lines and on dust) is left for future studies. 
We adopt a simple procedure for the calculation of ${\bf g_{\rm rad}}$: if the temperature of the dust, 
$T_{\rm d}>T_{\rm sub}$, where $T_{\rm sub}$ is the dust sublimation temperature (see further in the text) the force is assumed to be zero.

Our description takes into account two types of radiation: infrared and X-ray. 

The transfer of X-rays is treated in a single stream approximation. This implies that
X-rays are simply traced from the point source (i.e. corona) adopting  $e^{-\tau_{\rm X}}$ attenuation, where $\tau_{\rm X}$ is an optical depth with respect to X-ray absorption
\citep{Dorodnitsyn08b}. The distribution of the infrared radiation is calculated in a Flux Limited Diffusion approximation. 
Detailed discussion about the validity of equations (\ref{eq11})-(\ref{eq14}) is given in Paper II. 
We briefly remind that the closure relation between ${\bf F}$ and $E$ is obtained adopting the diffusion approximation:

\begin{equation}\label{FickLaw}
{\bf F}= - D \, \nabla E \mbox{,}
\end{equation}
where if optical depth is large $\tau\gg 1$, the diffusion coefficient reads:

\begin{equation}
D= c\, \lambda\mbox{,}\label{DifCoef}
\end{equation} 
where $\lambda=1/(\kappa \rho)$ is the photon mean free path, and $\kappa = \chi/\rho$.
From (\ref{radforce}), (\ref{FickLaw}), and (\ref{DifCoef}) one can see that
in the diffusion regime 
the radiation force does not explicitly depend on the opacity.

The diffusion approximation tacitly assumes that optical depth $\tau \gtrsim 1$.
Most of the torus where IR pressure is important indeed has $\tau_{\rm d}>1$ where 
$\tau_{\rm d}$ is the optical depth of the gas-dust mixture in the infrared.

If $\tau<1$ the diffusion approximation should be modified so a correct limiting behavior at
$\tau<1$ is regained. One can see that without such modification
when $\tau \ll 1$ the mean free path, $\lambda \to \infty$,  and
$D\to \infty$, and $|{\bf F}|\to \infty$ which is in contradiction with 
a free-streaming limit where it should be $|{\bf F}|\to c E$ (\ref{DifCoef}).
That is, when optical depth becomes small, or when $\rho\to 0$, the standard diffusion approximation is no longer applicable.

In order to describe correctly regions of small $\tau$, the standard approach is to adopt
the flux-limited diffusion approximation \citep{AlmeWilson74,Minerbo78,LevermorePomraning81}.
In this approximation $\lambda$ is replaced by $\lambda^{*} = \lambda\,\Lambda_{\rm F}$,
where $\Lambda_{\rm F}$ is the flux limiter. The flux limiter we adopted in Paper I,II and in the current work is that of \cite{LevermorePomraning81}:

\begin{equation}\label{LP_FluxLim}
\Lambda_{\rm F} = \frac{2+R_{\rm LP}}{6+3R_{\rm LP}+R_{\rm LP}^{2}}\mbox{,}
\end{equation}
where $R_{\rm LP}=\lambda\,|\nabla E |/E$. One can see, that if $\tau\to 0$, then $R_{\rm LP}\to \infty$, and 
$|F| \sim c\,E$; and if $\tau\gg1$ $R_{\rm LP}\to 0$ and $\Lambda_{\rm F}\to 1/3$.

\section{Interaction of radiation and the wind}
\subsection{Photoionization equilibrium}
The premise of the current work is
that the deposition of energy at parsec scale is dominated by external sources, i.e. by 
UV and X-rays from the central black hole and inner accretion flow. 
Due to the high opacity of the dust-gas mixture, a significant part of the UV and soft X-rays
is absorbed and reprocessed into infrared in a thin layer of thickness,
$\delta{\it l}/R_{1\rm pc} \simeq 1.3 \times 10^{-3}\,n_{7}^{-1}$ (Paper I).

The dynamical time within the flow is usually much larger than the characteristic time of the photoionization and recombination. Thus, the ionization balance is determined by the condition of photo-ionization equilibrium.
The condition of a photo-ionization equilibrium in the wind implies that 
the state of the gas 
can be be reasonably parameterized in terms of one ''ionization parameter'', $\xi$ -the ratio of radiation energy density to baryon density \citep{Tarter69}. In a popular form it reads

\begin{equation}\label{smallxi}
\xi=4\,\pi\,F_{\rm x}/n \simeq 4\cdot 10^2 \cdot f_{\rm x}\,\Gamma\, M_6/ (N_{23}\,r_{\rm pc})
\mbox{,}
\end{equation}
where $F_{\rm x}$ is the local X-ray  flux, $L_{\rm x}$ is the X-ray luminosity of the nucleus, and $n$ is the gas number density;
$N_{23}$ is the column density in $10^{23}$ ${\rm cm}^{-2}$, $f_{\rm x}$ is a fraction of the total accretion luminosity $L_{\rm BH}$ available in X-rays,  $\Gamma$ is a fraction of the total 
Eddington luminosity $L_{\rm edd}=1.25\cdot 10^{44}\,M_6$, where $M_{6}$ is the mass of a BH in $10^{6}M_{\odot}$.  We adopt a simple attenuation model in which the local deposition of energy is proportional to $F_{\rm x}=L_{\rm x}e^{-\tau_{\rm x}}/(4\pi r^2)$, where the optical depth at soft X-rays is integrated over a straight line connecting the point source corona with a given parcel of gas
${\tau_{\rm x} = \displaystyle \int \, \kappa_{\rm x}\rho \,dl }$, and  $\kappa_{\rm x}$ is the X-ray opacity.

\subsection{Hot and cold gas}
To calculate photoionization balance, 
we take into account Compton and photo-ionization heating and Compton, radiative recombination, bremsstrahlung and line cooling.  The rates of these processes are calculated making use of the XSTAR photo-ionization code
\citep{KallmanBautista01}
for the incident spectrum which is a power law with energy index $\alpha$. 
As the direct incorporation of the XSTAR subroutines into the hydrodynamics code is not feasible, the key ingredient of our method is to make use of 
approximate analytic formulas for the heating and cooling rates. 
These formulae approximate extensive tables of heating-cooling rates obtained from XSTAR; 
these approximations are
similar to those of \cite{Blondin94}, although modified by \citet{Dorodnitsyn08b}
to incorporate better atomic data and power law ionizing continuum with $\alpha$=1:

\begin{equation}\label{heating1}
\Gamma_{\rm IC}({\rm erg\,cm^{3}\, s^{-1}})= 8.9\cdot 10^{-36}\,\xi\,(T_{\rm x}-4T)\mbox{,}
\end{equation}
for the Compton heating - cooling;

\begin{equation}\label{heating2}
\Gamma_{\rm x}({\rm erg\,cm^{3}\, s^{-1}})= 1.5\cdot 10^{-21}\,\xi^{1/4}\,T^{-1/2}(T_{\rm x}-T)T_{\rm x}^{-1}\mbox{,}
\end{equation}
for the photo-ionization heating-recombination cooling, and
for the bremsstrahlung and line cooling:

\begin{eqnarray}\label{heating3}
\Lambda({\rm erg\,cm^{3}\, s^{-1}})&=&3.3\cdot 10^{-27} T^{1/2}\nonumber\\
&+&(4.6\cdot 10^{-17} \exp(-1.3\cdot 10^5/T)\xi^{(-0.8-0.98\alpha)} T{-1/2}+ 10^{-24})\, 
\delta\mbox{.}
\end{eqnarray}
Formulae (\ref{heating1})-(\ref{heating3}) 
were found to be in reasonable (25\%) agreement with numerical results.

A variety of physical and chemical processes influence the transformation of X-rays into IR if dense dusty, molecular gas is illuminated by 
external X-rays \citep{Maloney96,KrolikLepp89}. 
Rather than attempt to incorporate all these processes in our radiation hydrodynamics framework, 
in the current paper we adopt several approximations which reproduce the  qualitative behavior of low temperature gas exposed to
external X-ray radiation.
At column densities, $N$ smaller than $10^{24}\, {\rm cm^{-2}}$ photons with energies below
$10$ keV interact with the gas through photoelectric absorption while Compton scattering dominates at higher energies. The total ionization per particle at a 
given point in the wind
is proportional to $\xi_{\rm eff}$, the ionization parameter which takes into account the frequency dependent attenuation of X-ray flux.
It is shown by \cite{Maloney96} that heating and cooling rates of such X-ray illuminated molecular gas depend on the modified ionization parameter:
$\xi_{\rm eff} =\xi/N_{22}^{0.9} = 1.26\times 10^{-1} F_{\rm x}/ (n_{8} N_{22}^{0.9})$, and several regimes are approximately distinguished:
1)
The regime of highly ionized gas $\xi_{\rm eff}\gg1$.  In this regime for the heating and cooling rates we make use of approximations 
(\ref{heating1}) - (\ref{heating3});
2)
If $ 10^{-3} \ll  \xi_{\rm eff} \lesssim 1$ the gas is primarily atomic, and we also adopt heating-cooling rates (\ref{heating1}) - (\ref{heating3});
3)
If the ionization parameter, is small 
$(\xi_{\rm eff} \ll  \xi_{\rm m} =10^{-3})$ the gas is largely molecular. This is where the most complexities due to chemistry and radiative interaction with the dust occur. 
The proper incorporation of this regime calls for consideration of all the radiative, chemical and dust network of processes. This is challenging even without coupling to hydrodynamics. 

In the regime when $\xi_{\rm eff} <  \xi_{\rm m}$, 
we assume that the molecular gas is in radiative equilibrium with dust, i.e. $T_{\rm g} = T_{\rm d}$ where equilibrium dust temperature is found from the attenuated X-ray flux, $F_{\rm x}$.
Thus we approximately take into account that
dust can directly reprocess X-rays to IR contributing to the IR energy density, $E$. We approximately take this contribution into account, using equation (see, Appendix B):

\begin{equation}\label{dustXupdate}
\frac{d T_{\rm r}}{dt}=n_{\rm d}\,c\, \sigma_{d}\left({ T_{\rm d}} - { T_{\rm r}}\right)\mbox{,}
\end{equation}
Dust temperature is found from the approximate relation, 
$T_{\rm d}= (4F_{\rm X}/(ac))^{1/4}$ (notice that $E=aT^{4}_{\rm r}$), where $F_{\rm X}$ is the local {\it attenuated }X-ray flux. If the dust temperature $T_{\rm d}>T_{\rm sub}$, where $T_{\rm sub}$ is the dust sublimation temperature then it is assumed that dust is destroyed and no conversion to IR occurs.

In our approach,
the main difficulty in calculating the energy density of IR radiation from X-rays resulting from a single fluid approximation. By having only one species (gas) and not evolving a separate dust component, we can mimic the interaction between dust and X-rays, and calculate $T_{\rm d}$ by means of  equation
(\ref{dustXupdate}), while additionally assuming the relation between $T_{\rm d}$, and $T$ in a dusty-molecular phase, in which, for simplicity we assume $T=T_r$, with the later calculated from(\ref{dustXupdate}).
Other dust characteristics in (\ref{dustXupdate}) are as follows:
$n_{\rm d}=\rho f_{\rm d}/m_{\rm d}$ is the dust grain number density, where $f_{\rm d}$ is the
dust-to-gas mass ratio, and $m_{\rm d}$ is the dust grain mass, and
$\sigma_{\rm d}=\pi r_{\rm d}^{2}$ is the dust cross-section, and $r_{\rm d}$ is the dust grain radius.
Throughout our calculations, we adopt the following fiducial 
values for the parameters of dust: typical dust grain radius, $r_{\rm d}=1\times 10^{-4}\,{\rm cm}$,
the density of the dust grain $\rho_{\rm d}=1\, {\rm g\, cm^{-3}}$, $f_{\rm d}=1\times 10^{-2}$, and above the sublimation temperature $T_{\rm sub}=1500\,{\rm K}$ the dust is assumed to be destroyed.
Equation (\ref{dustXupdate}) for $T_{\rm d}(F_{\rm X})$ approximately describes how X-rays are converted into IR through reprocessing by dust.

If $\xi_{\rm eff} \geqslant \xi_{\rm m}$ and the $T<T_{\rm sub}$ we assume that the exchange between $T_{\rm r}(E)$ and $T(e)$ is described by the balance between $D_{t}(E/\rho)$ and the last two terms in equation (\ref{eq14}). The gas temperature $T$ can contribute to IR and
vice versa. For example, gas can contribute to the X-ray-IR conversion through cooling. Cooling can be due to 
radiative processes, i.e. the low energy limit of (\ref{heating2})-(\ref{heating3}) and due to ''adiabatic'' cooling due to fluid motion.

The update of $E$ from (\ref{dustXupdate})
is performed adopting operator splitting in the source step, i.e., before the update of the equation (\ref{eq14}) is done. 
During the update over $dt$, the new $E_{\rm r}(t+dt)$ is calculated from (\ref{dustXupdate}).

\section{Methods}

To solve the system of equations (\ref{eq11})-(\ref{eq14}) we expanded the radiation hydrodynamic framework described in our previous papers.
The additions include the interaction of X-rays with dust grains and with gas. 
The coupling between the X-ray heating and cooling with hydrodynamics has 
been implemented and tested in our earlier works \citep{Dorodnitsyn08b}.
Here we have added physics which includes the direct interaction of X-rays with dust grains. 

The radiation-hydrodynamic equations are solved using the ZEUS code \citep{Dorodnitsyn12a} with extensions taken from \cite{Dorodnitsyn08b}. 
The radiation-hydrodynamic part is designed and tested in \citep{Dorodnitsyn11a, Dorodnitsyn12a} and conforms with methods and structure of the original ZEUS code \citep{Stone92a}.

In the following we briefly review the organizational structure of the code.
Similarly to the way pure hydrodynamics is treated in ZEUS, the radiation 
part is also split into the source and transport steps. Artificial viscosity is adopted to resolve discontinuities in the flow.
In the source step the difference equations for $\partial E/\partial t$ and $\partial e/\partial t$ are solved, and in the transport step the advection of $E$ and $e$  is calculated. 
The update of $E$ and $e$ is calculated adopting fully implicit scheme (see Paper II); this includes  
the solution of the diffusion problem and the $e\leftrightarrow E$ update i.e. the update of gas and radiation energy densities due to radiation-gas  interaction.

Similarly, in the X-ray-matter interaction step, heating $\gamma_{\rm x} \rightarrow e $, and cooling $e\rightarrow \gamma_{\rm x}$ due to Compton, radiative recombination, bremsstrahlung and line cooling
is also done fully implicitly in a separate update.

We make use of the ${R,z}$ cylindrical grid, which extends from $R_{\rm in}$ to $R_{\rm out}$ in radial and from $z_{\rm in}=0$ to $z_{\rm out}=R_{\rm out}$ in the vertical direction. 
The lower boundary at $z_{\rm in}=0$ is intended to represent a thin accretion disk, which 
serves as an infinite reservoir of gas and dust.  Wind can escape from all other boundaries, 
in principle, though the centrifugal barrier effectively prevents escape from the 
inner boundary.
A fixed grid of size $200\times200$ is adopted for all calculations.
Hydrodynamic
boundary conditions (BC) are as follows: at the left (innermost), right, and upper boundaries we adopt outflowing boundary conditions. 

At the equator we fix the inflow speed $v_{0, j}$ at a small fraction of the speed of sound leaving the density to adjust according 
to the continuity equation.
Hereafter, index $j$ varies over $R$ and index $i$ over $z$.
We allow $\rho_{is,j}$ to vary and fix $v_{is,j}$, that is 
$\rho_{is,j}=v_{is+1,j} \rho_{is+1,j}/v_{is}$, and $v_{is}=const$, provided $v_{is}\ll v_{is+1, {\rm s}}$,
where $v_{is+1, {\rm s}}$ is the sound speed; 
everywhere in the simulations we assumed $v_{is} / v_{is+1, {\rm s}} =0.01$. 
Notice that this is different from the implementation of the equatorial BC in Paper II where it was assumed $\rho_{is,j} \sim R_{j}^{-p}$ and velocity was found from the continuity equation and only checked after the fact
that $v_{is,j}\ll v_{is,\rm s}$.
Present implementation of boundary conditions allows the disk to be flexible choosing mass loading of the wind. The initial distribution of density at the equator was taken to be $\rho_{0}r_{is}^{-1}$, where $\rho_{0}=m_{p} n_{0}$ is the characteristic density, $m_{p}$ is a proton mass. 
We find that numerical solution quickly forgets about initial equatorial distribution of $
\rho$, and that solutions are not sensitive to the choice of $p$.
Theoretically, since $\rho_{is,j}$ is allowed to change, the final solution should only weakly depend on initial equatorial distribution
of $\rho$. 
Practically, however, this is true only for those parts of the disk where the wind is developed. 
We also find that the present implementation of BC produces a more stable solution near the equator, thus somewhat relaxing the 
restriction on the hydrodynamical time step. 

The radiative BC are free-streaming $F\sim cE$ at all boundaries except at the equator where zero flux BC are implemented.

\section{Results}
\subsection{Dynamics of the wind}

Deposition of radiation energy into the flow is controlled by the parameter $\eta=L_{\rm bol}/L_{\rm Edd}$. 
We calculate models for $\eta=\{0.05, 0.1, 0.3, 0.4, 0.5,0.6\}$. 
{\mybf The innermost boundary of the computational domain is located at radius, $R_{0}=0.5\,\text{pc}$.}
Other parameters
are $M_{\rm BH}=1\times 10^{7} M_{\odot}$, $f_{\rm X}=0.5$, and the initial scale density $ n_{0}=1 \times 10^{8} \, {\rm cm^{-3} } $. The evolution of the radiative flow was followed for 10-30 dynamical times,
$t_{0}\simeq 1.5\times 10^{11}\, r_{\rm pc}^{3/2} M_{7}^{-1/2}$s. 

In Papers I and II the temperature of the IR radiation was prescribed at the boundary, in a physically motivated way (to match the energetics of X-rays, $f_{\rm x} L_{\rm bol}$). 
The results obtained in Paper II are generally in a good agreement with the present modeling. However, explicitly
including X-rays into calculations reveals several important features of the flow which 
were absent in previous studies.

One of the results of the present study is that 
the pressure of the hot photo-ionized component is very important in shaping the IR flow.
Excitation of such a hot, photoionized wind automatically mirrors the creation of the IR flow.
The interface between the two components of the  flow is seen in all simulations presented in this paper.
In all cases it takes the form of  oblique shocks or contact discontinuities. 
A supersonically moving hot gas obliquely hits the slowly moving IR flow creating a discontinuity (in most cases an
oblique shock wave).  

\subsubsection{Model with $L = 0.6 L_{\rm edd} $ }
Figure \ref{figGam06} shows the development of the radiation-driven disk wind for $\Gamma=0.6$.
The outflow develops quickly after about one dynamical time, $t_{0}\simeq 4\times 10^{4} \rm yr$.
The mass-loss rate  is $\langle {\dot M} \rangle  \simeq 0.1-0.2 M_{\odot}\, {\rm yr}^{-1}$ on average, and  peaks at $\langle {\dot M} \rangle  \simeq 1 M_{\odot}\, {\rm yr}^{-1}$ in the period of time between 
$4\times 10^{3}$ yr and $5\times 10^{3}$ yr.  The total computation spans from 0 to $8\times 10^{3}$yr. These averaged peak numbers include both failed wind and wind which has enough energy to escape to infinity.

The energetics of the wind is measured by computing the 
kinetic output of the wind at the outer boundary of the computational domain $\Sigma$:
$v_{\rm kin} \simeq 2 L_{\rm kin}/{\dot M}$, where $\displaystyle L_{\rm kin}=\int_{\Sigma} \rho v^{3}/2\, d\Sigma$ is the kinetic luminosity of the wind.  For a radiation-driven wind we expect low values of 
$L_{\rm kin}/L_{\rm bol}$; here we have $L_{\rm kin}/L_{\rm bol} \simeq 5\times 10^{-5}-2 \times 10^{-4}$.

The averaged dynamics of the wind,  is described by the 
the average bulk velocity of the flow $\displaystyle \langle v \rangle = {\int_{V}\,\rho v dV} / {\int_{V}\, \rho dV}$, where $V$ is the total or partial volume occupied by the flow. Here, the dense part of the flow has 
$ \langle v_{z} \rangle \simeq \text{ a few} \times  1{\rm km\, s^{-1}}$, and $\langle v_{R}\rangle\simeq \text{ a few} \times 10\,{\rm km\, s^{-1}}$.  The maximum velocity reaches $1000\, {\rm km\, s^{-1}}$ in $R$, and over 
$600 \, {\rm km\, s^{-1}}$ in the $z$ direction.

\begin{figure}[htp]
\includegraphics[width=600pt]{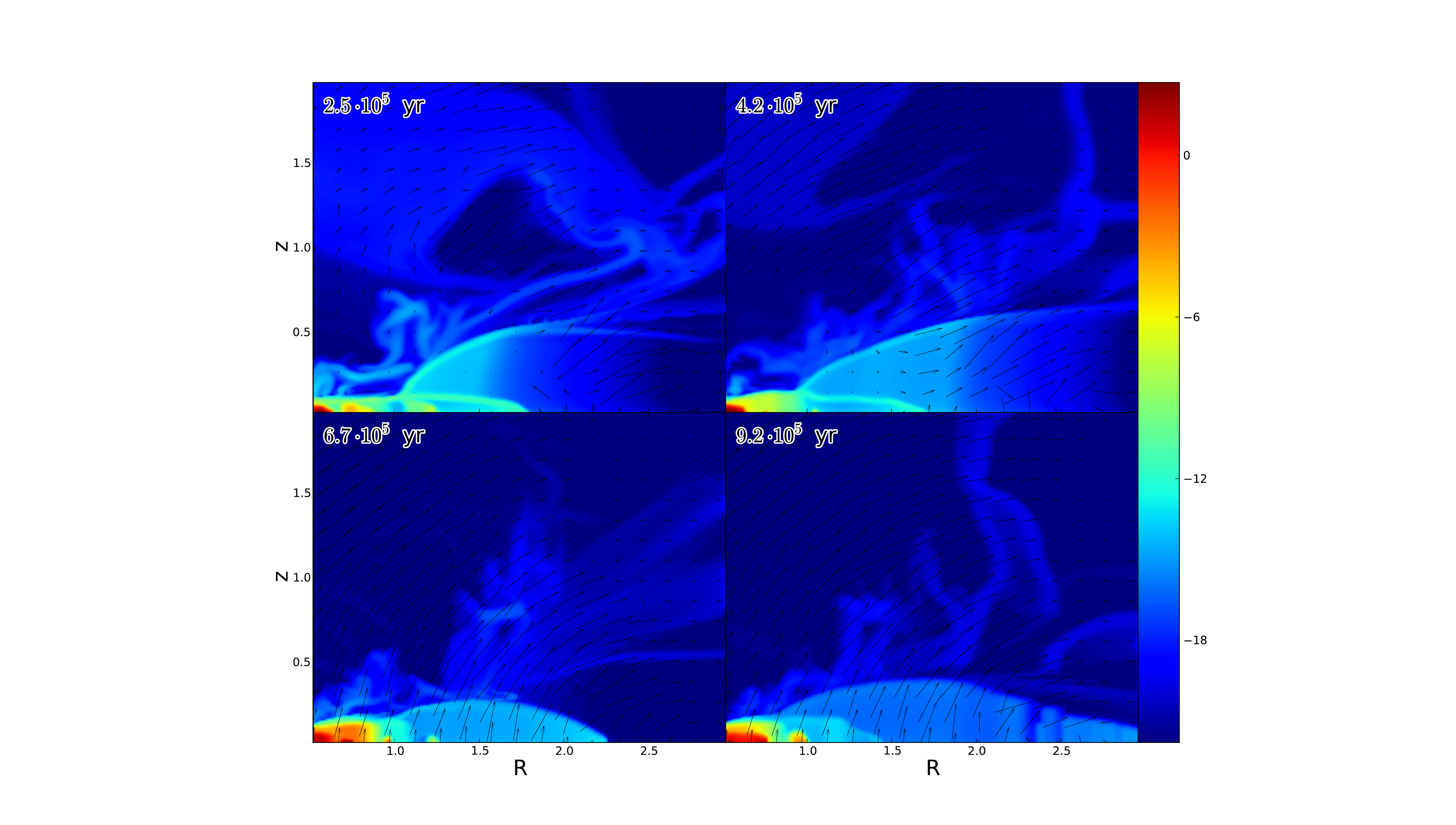}
\caption{
Color plot of the density, $\log\rho$ in ($g\,{\rm cm^{-3}}$) for the model with $L= 0.6 L_{\rm edd}$
shown at different times given in years.
Axes: horizontal: $z$: distance from equatorial plane in parsecs; $R$: distance from the BH in parsecs;
}\label{figGam06}
\end{figure}

Figure \ref{figTgas} shows a color plot of the distribution of the gas temperature. 
The gas spans a vast range of temperatures from being in a cold molecular state to 
photo-ionized, high temperature warm absorber flow. 
An apparent feature of our models is that the hot wind consists of large scale inhomogeneities.
The cold flow doesn't show such large scale structure. 
Notice that we have a very simple test if the gas is in the cold, molecular-dusty phase,  
$\xi_{\rm eff} <  \xi_{\rm m}$, and that if hot component is not shown the density structure is much more pronounced in the temperature plot.

{\mybf It is of interest to address the question of whether there can be the cold gas in a hot wind as well.
However, the limited resolution of our studies does not allow us to provide a reliable answer to this question.  We do not find the coexistence of hot and cold phases of gas in any appreciable quantities, but this does not exclude the possibility of such coexistence on length scales smaller than we can resolve.
}

\subsubsection{Model with $L = 0.5 L_{\rm edd} $: the importance of the hot flow pressure}

Figure \ref{figGam05} which shows a color plot of the density at different times shares a lot of similarity with 
Figure \ref{figGam06}. One can distinguish approximately three regions in the flow: 
1) a disk-like, geometrically thin and dense region, extending to $R \sim 1.5\,$pc; 
2) the IR-driven flow with lower density, separated by contact discontinuity from 
3) a hot, photoionized flow. 
As $L$ decreases, so does the energetics of the wind. However, the vertical thickness of the dense IR supported flow is determined by the delicate balance between a) the $\rho v^{2}$ pressure of the hot component from above, and b) by the pressure of IR supported wind from below.

Decreasing $\Gamma$ from 0.6 to 0.5 causes the pressure of the hot flow to be less than in the solution shown in Figure \ref{figGam05}. 
The balance between the pressure of the photo-evaporated wind and IR drive dusty flow is 
is slightly shifted towards IR-dominated flow.

Figure \ref{figGam05Velos} shows the 2D distributions of the velocity components $v_{z}$ and $v_{R}$. 
The slowly outflowing wind is clearly seen; its well defined boundary extends to $~0.5\,$pc.
The flow consists of a ''core'', i.e. slow moving wind with radial velocities of a 
few$\times 10\lesssim 100\,\rm km\,s^{-1}$, and $v_{z}$ much smaller, of the order of the a few$\times 1\,\rm km\,s^{-1}$.  
Fast lower density component of the flow occupies most of the domain having velocities of the order of a few$\times 100\lesssim 700\,\rm km\,s^{-1}$.
The overall characteristics of the velocity field are similar to those of the model with $L = 0.6 L_{\rm edd} $.
Not surprisingly, pumping less energy initially in X-rays generate less powerful wind:
$L_{\rm kin}/L_{\rm bol} \simeq 1\times 10^{-5}-7 \times 10^{-5}$.

The mass-loss rate $\langle {\dot M} \rangle  \simeq 0.1 M_{\odot}\, {\rm yr}^{-1}$ with the peak $\langle {\dot M} \rangle  \simeq 1.1 M_{\odot}\, {\rm yr}^{-1}$ in the period of time between 
$8\times 10^{3}$ yr and $10\times 10^{3}$ yr, and the computation spans from 0 to $8\times 10^{3}$yr. 

\begin{figure}[htp]
\includegraphics[width=400pt]{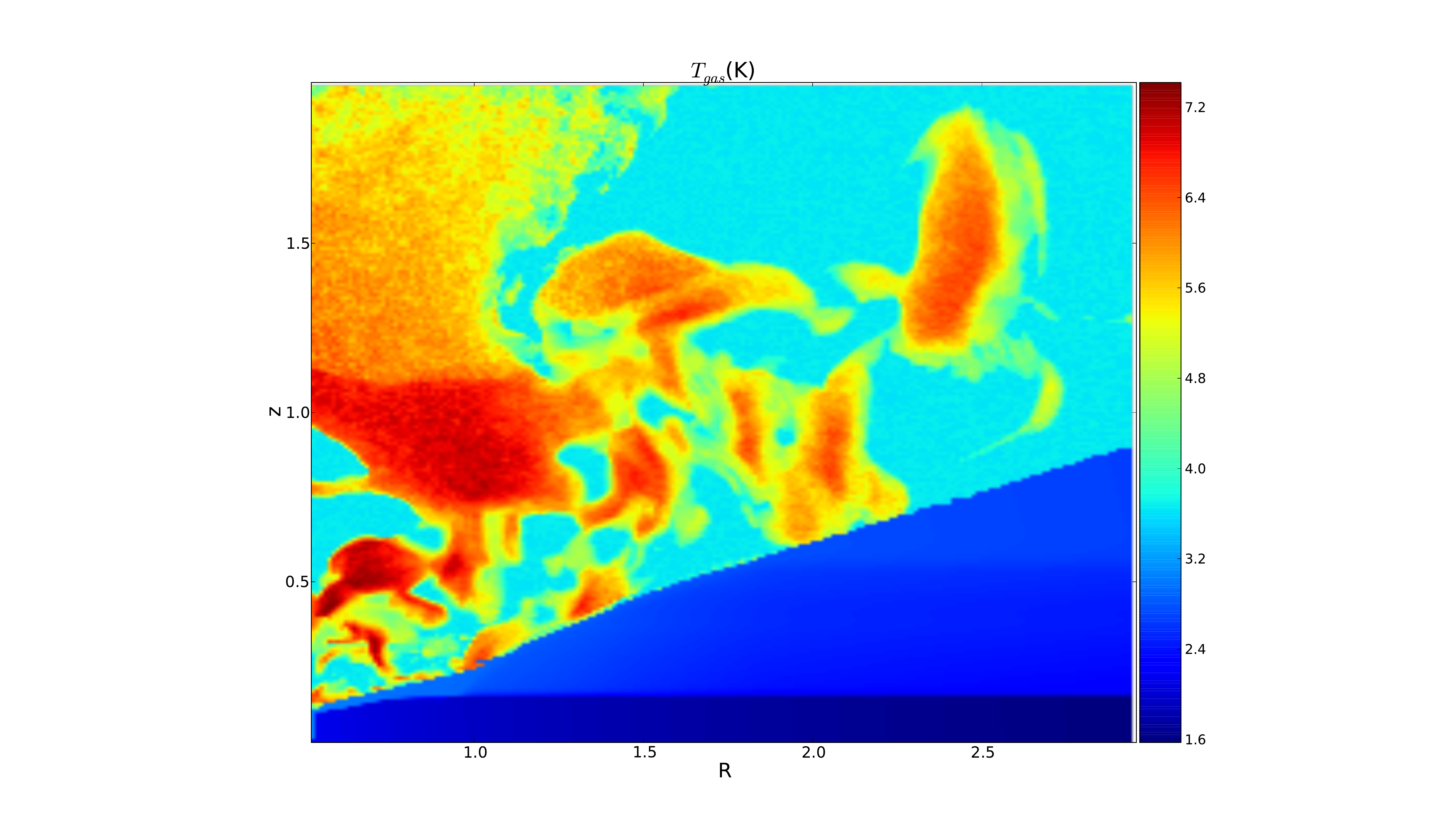}
\caption{
Color plot of the gas temperature, $\log T$ in K for the model with $L= 0.6 L_{\rm edd}$,
after $4.2 \cdot 10^5$ yr.
Axes: horizontal: $z$: distance from equatorial plane in parsecs; $R$: distance from the BH in parsecs;
}\label{figTgas}
\end{figure}

\begin{figure}[htp]
\includegraphics[width=600pt]{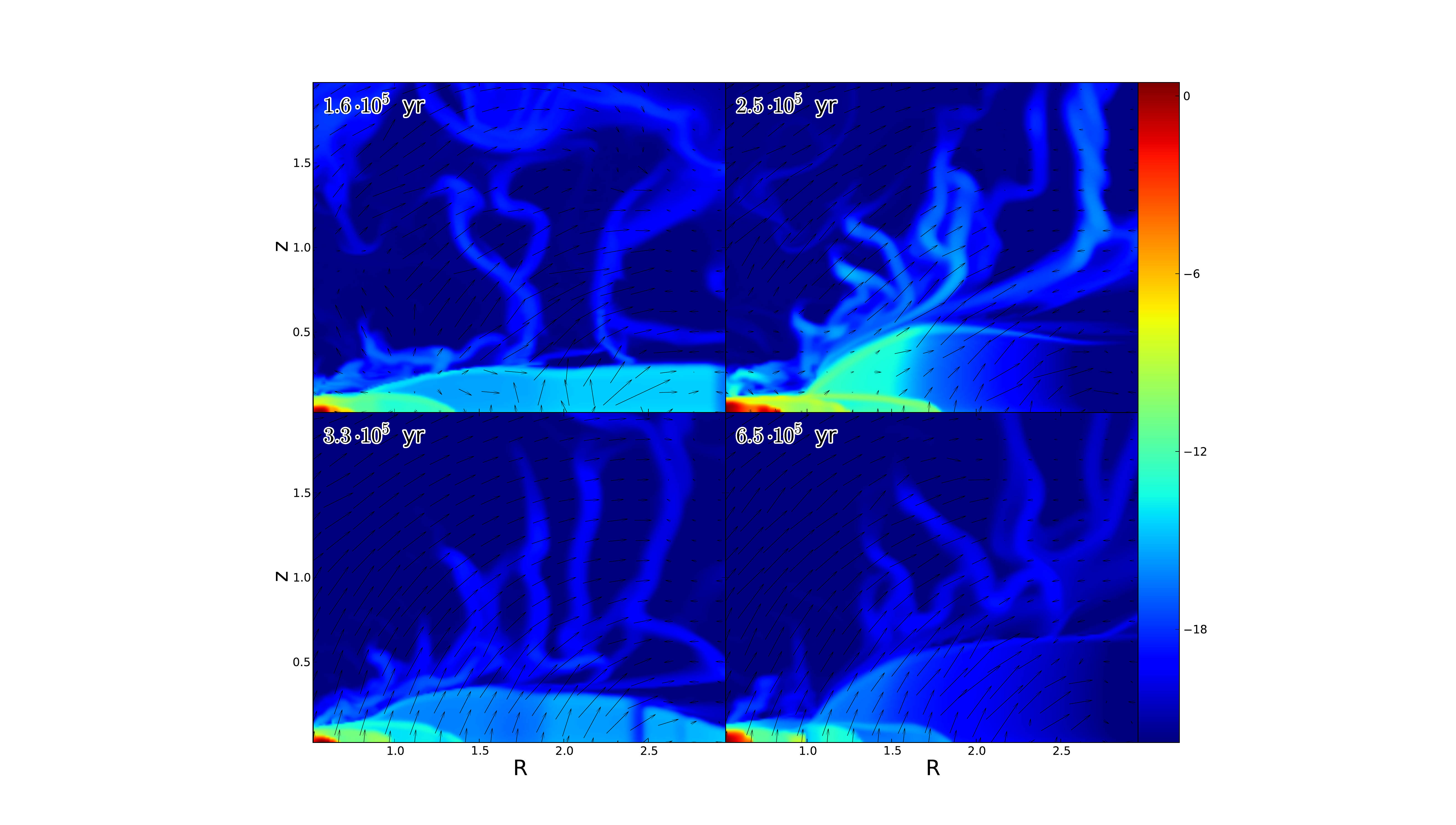}
\caption{
Color plot of the density, $\log\rho$ in ($g\,{\rm cm^{-3}}$) for the model with $L= 0.5 L_{\rm edd}$,
shown at different times given in years.
Axes: horizontal: $z$: distance from equatorial plane in parsecs; $R$: distance from the BH in parsecs;
}\label{figGam05}
\end{figure}

\begin{figure}[htp]
\includegraphics[width=600pt]{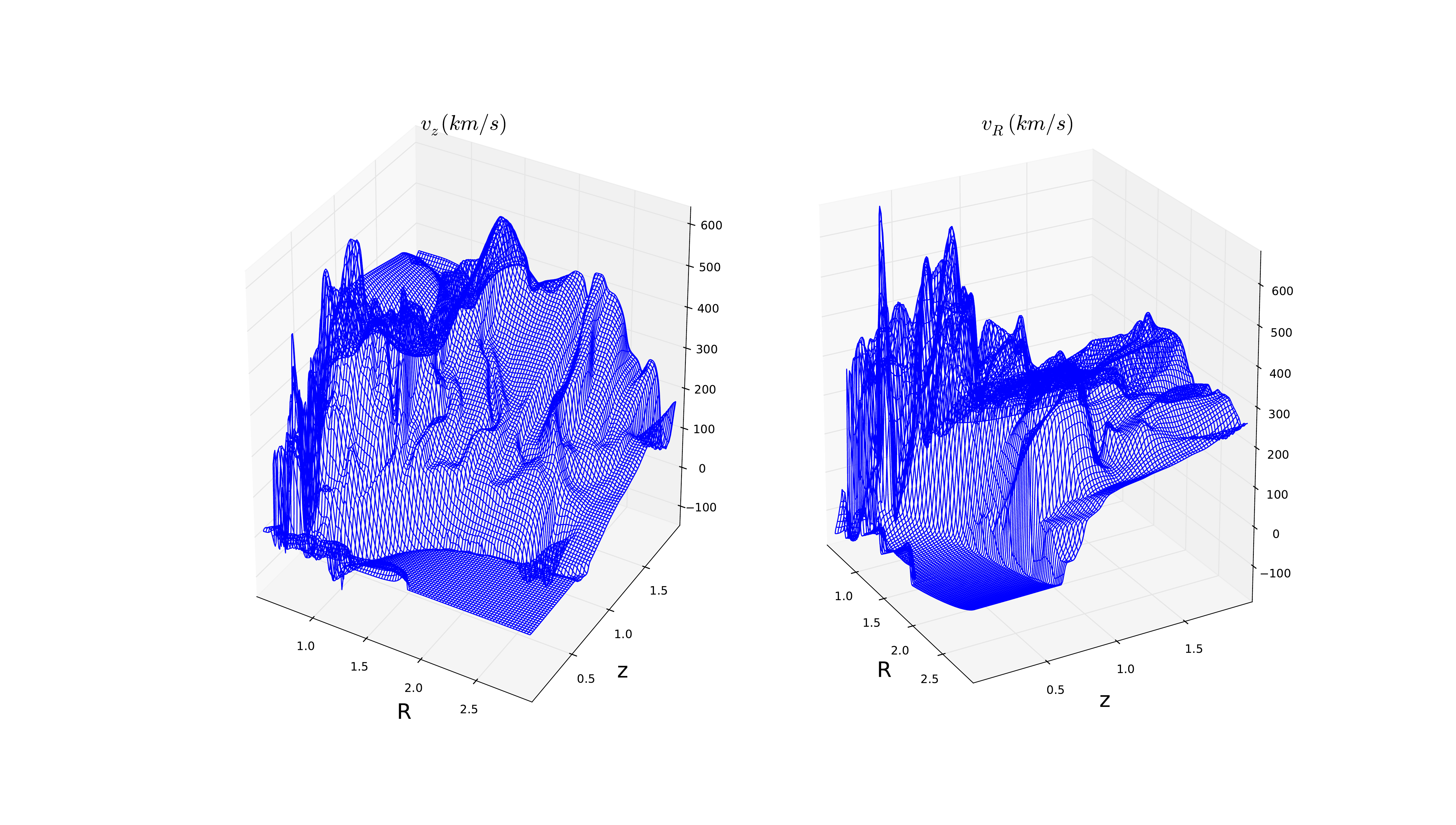}
\caption{The surface plot of the $z-$ and $R-$ velocity components. 
Model is shown with $L= 0.5 L_{\rm edd}$, $t=6.5\times 10^{5}\,{\rm yr}$;
Horizontal: $R$: distance from the BH in parsecs;
$z$: distance from the equatorial plane in parsecs;
}\label{figGam05Velos}
\end{figure}

\subsubsection{Model with $L = 0.3 L_{\rm edd} $}
Figure \ref{figGam03} demonstrates the evolution of the density for this model at different times. The disk-wind system goes through several stages: the domination of the hot wind; puffing up the IR supported 
disk; squeezing of the later towards the equator by the hot wind; building a high density disk-like outflow, etc. As in previous examples, the system doesn't come to a quasi-stationary state instead going through such episodes all over again.

An interesting feature which is seen at $t=2.5\times 10^{5}\,{\rm yr}$ is in fact the residue from the contact discontinuity which existed at earlier times.
The average velocity
$\langle v \rangle \simeq \text{ a few} \times 172\,{\rm km\, s^{-1}}$. This is about $60\%$ of the 
escape velocity, $U_{\rm esc}$; the numbers are given for $t=20\, t_{\rm dyn}$. The maximum velocity of the dusty flow is $242\,{\rm km\, s^{-1}}$.

The value of the mass-loss rate is remarkably similar to models with higher $\Gamma$: $\langle {\dot M} \rangle  \simeq 0.1-0.2 M_{\odot}\, {\rm yr}^{-1}$ reaching $1M_{\odot}\, {\rm yr}^{-1}$. 

\subsubsection{Models with $L \lesssim 0.1 L_{\rm edd} $: slow accretion with episodic outflow}
When the radiation input falls below 
an approximate threshold value no
rigorous outflow is observed in our simulations. Instead we see that episodic outbursts of 
the hot evaporative flow are excited from the inner parts of the disk. 
Here we found this threshold luminosity to be $L \simeq 0.1 L_{\rm edd}$. The results for the model
for $L = 0.1 L_{\rm edd}$ is shown in Figure \ref{figGam01}.

The density of this wind is too low to provide any considerable shielding from X-rays. 
Without much shielding it is difficult to form cold and dusty phase which can be accelerated by IR. 
Such an episodic outburst of the dense wind is followed by a gradual fall back of the dense component mostly in $z$ direction towards equatorial disk. 

The velocity of the hot wind, $v<U_{\rm esc}$, and it 
skirts the denser flow and falls back towards equator at larger radii, $R\simeq 2.5$pc. The density piles up there and 
the shielding from X-rays rapidly increases. Correspondingly, the ionization parameter drops
below $\xi_{\rm m}$ and the cold IR-driven wind again has a chance to develop. 
Notice that the vertical component of the radiation force is roughly 
${\bf g}_{{\rm eff},z}\sim (dE/dr) \cos\theta$, where $r$ is the spherical radius, and $\theta$ is the inclination angle, measured from $z$ axis.
Since the vertical extent of the IR wind is small, ${\bf g}_{{\rm eff},z}$ is small, and 
the dynamic pressure of the IR wind cannot overwhelm the pressure of the hot component at higher $z$.
The IR wind stalls and starts to slowly fall back (accrete) towards the equator.
Density outside this dense IR-supported ''pancake'' drops and the gas gets overionized again. The cold disk itself shrinks from both inside and outside.

We also calculated a model with $L = 0.05 L_{\rm edd}$, and found that the general behavior of this
model is similar to previous with $L = 0.1 L_{\rm edd}$. The wind is episodic and intermittent with 
the ''IR-supported pancake'' phase.

It is difficult to access the dynamical properties of models with $L\lesssim 0.1\, L_{\rm edd}$, because our equatorial BC excludes the influx of matter into the accretion disk. 
Strictly speaking by assuming thin accretion disk as a source of matter for the wind together with outflowing BC conditions we assumed that the wind, even if very weak, is always present. We will further elaborate on this limitation in the following section.

\begin{figure}[htp]
\includegraphics[width=600pt]{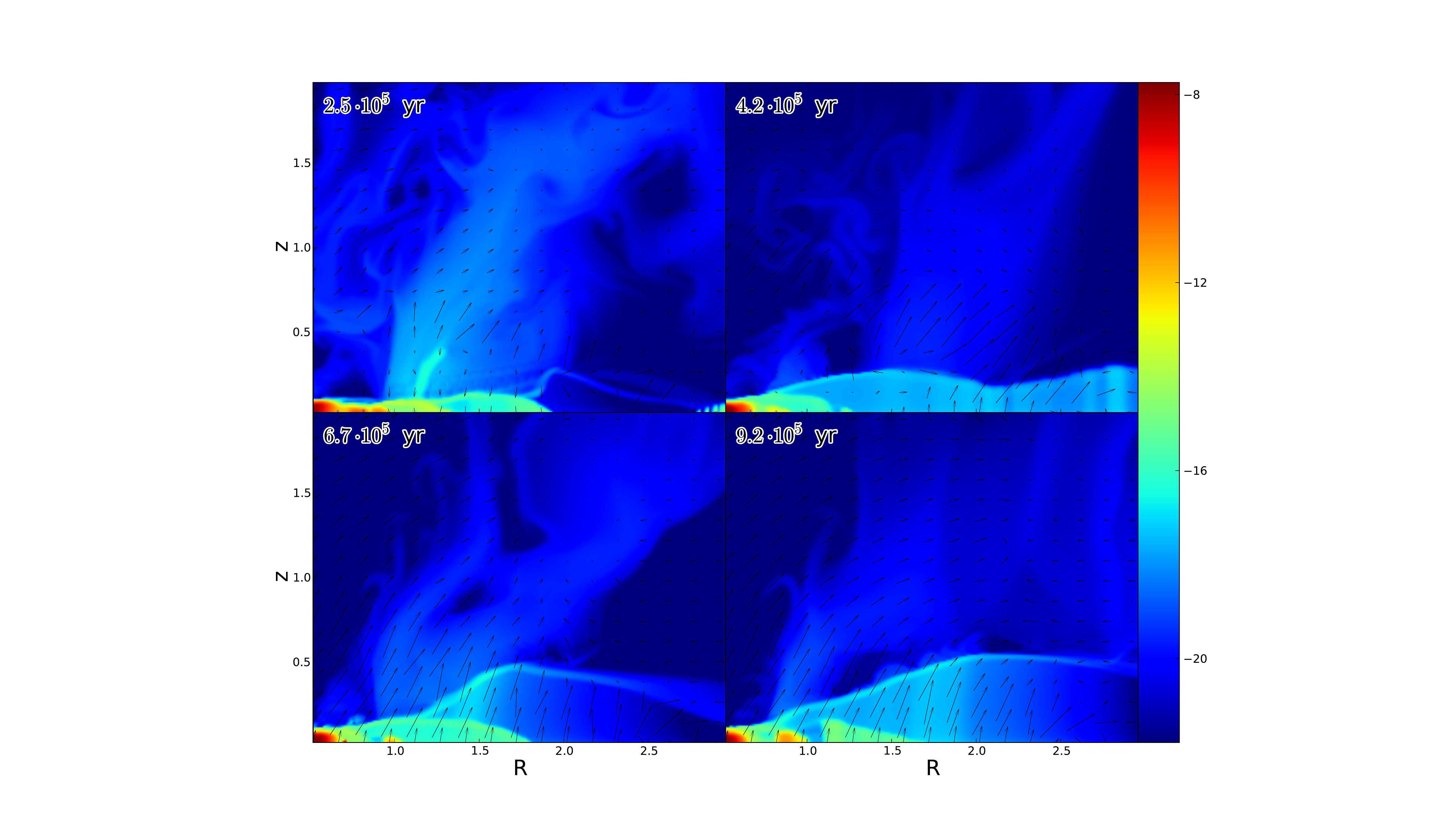}
\caption{
Color plot of the density, $\log{\rho}$ in ($g\,{\rm cm^{-3}}$) for the model with $L= 0.3 L_{\rm edd}$
shown at different times given in years.
Axes: horizontal: $z$: distance from equatorial plane in parsecs; $R$: distance from the BH in parsecs;
}\label{figGam03}
\end{figure}

\begin{figure}[htp]
\includegraphics[width=600pt]{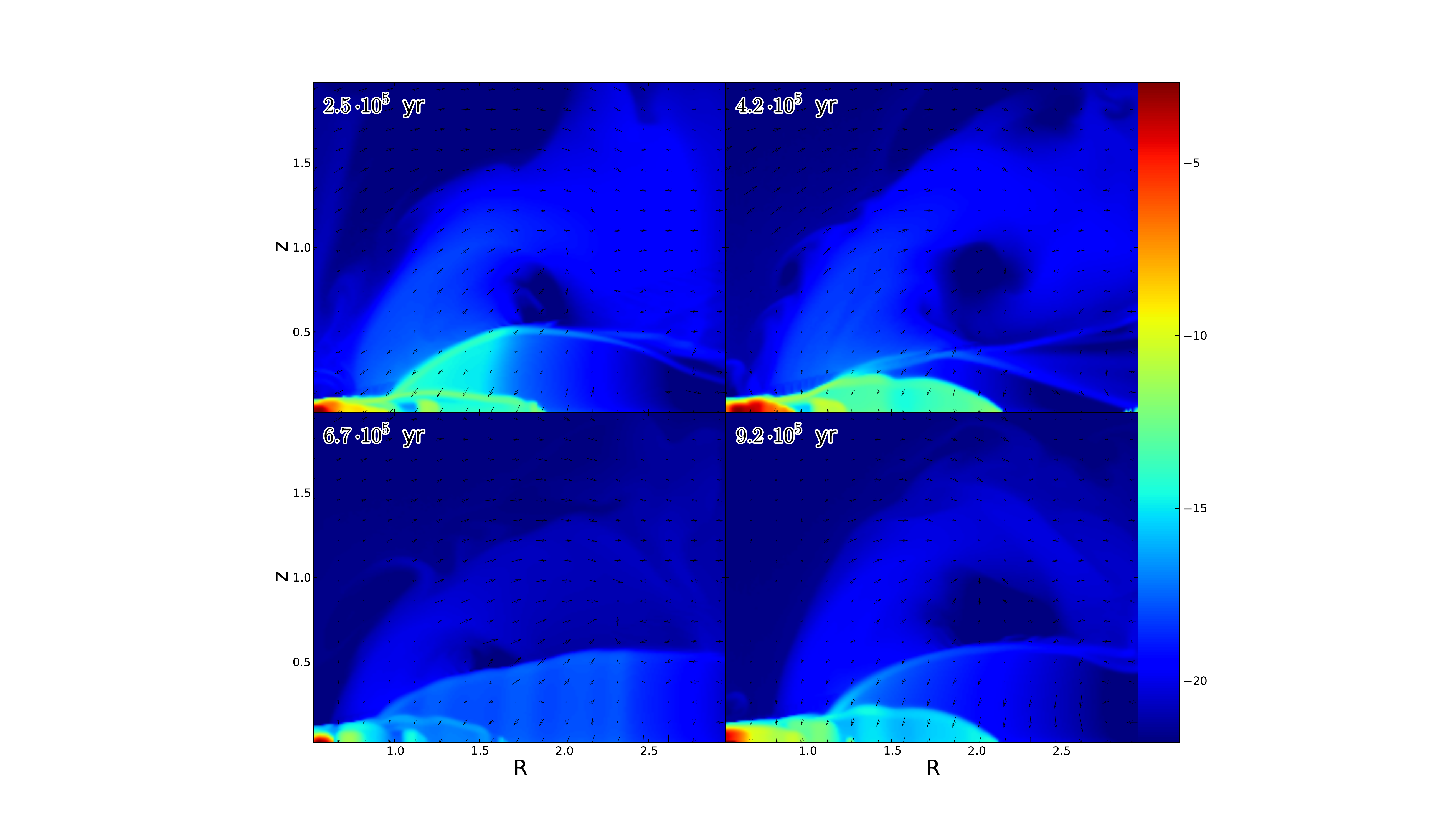}
\caption{
Color plot of the density, $\log{\rho}$ in ($g\,{\rm cm^{-3}}$) for the model with $L= 0.1 L_{\rm edd}$
shown at different times given in years.
Axes: horizontal: $z$: distance from equatorial plane in parsecs; $R$: distance from the BH in parsecs;
}\label{figGam01}
\end{figure}

\section{Obscuring properties}
We calculate 
Thomson optical depth, $\tau_{\theta}=  \int_{l} \, \kappa_{e} \rho \, dl $, where $l$ is measured along the line of sight at the angle, $\theta$ from the vertical axis, and averaging is made over all calculated models for particular $\Gamma$. Processing different models, we are interested in an angle $\theta_{\rm ph}$ where $\tau_{\rm ph}= \tau({\theta_{\rm ph}})=1$, i.e. when the wind becomes opaque. 

In the following we summarize results, listing models in the order from low to high luminosities.
When $\Gamma = 0.05$ the optical depth rises sharply from 0.1 to 0.5 at 
$\langle \theta \rangle \simeq 75^{\circ}$, (hereafter $\theta_{\rm edge} $) and the photosphere is at $\langle \theta_{\rm ph} \rangle\simeq 78^{\circ}$.

Increasing the luminosity to $\Gamma = 0.1$ causes the dense part of the wind to become
thicker in vertical extent,
optical depth rises at the angle
$\theta_{\rm edge} \simeq 73^{\circ}$, and the photosphere is at $\langle \theta_{\rm ph} \rangle\simeq 75^{\circ}$.
At $\Gamma = 0.3$ optical depth rises sharply from 0.1 at 
$\langle \theta \rangle \simeq 73^{\circ}$, $\langle \theta_{\rm ph} \rangle\simeq 75^{\circ}$.
and $\theta_{\rm edge}\simeq 73^{\circ}$.

For the model with $\Gamma = 0.5$ 
the photosphere is at $\langle \theta_{\rm ph} \rangle\simeq 75^{\circ}$ and the torus edge is at
$\theta_{\rm edge} \simeq 73^{\circ}$.

Figure \ref{figtau} shows a color plot of a radial Thomson optical depth measured from a BH to a given point of the computational domain. The model shown has $L= 0.6 L_{\rm edd}$.
This is the most luminous model we have calculated. It has
the photosphere located at $\langle \theta_{\rm ph} \rangle\simeq 72^{\circ}$ and the torus edge is at
$\theta_{\rm edge} \simeq 71^{\circ}$.

\begin{figure}[htp]
\includegraphics[width=400pt]{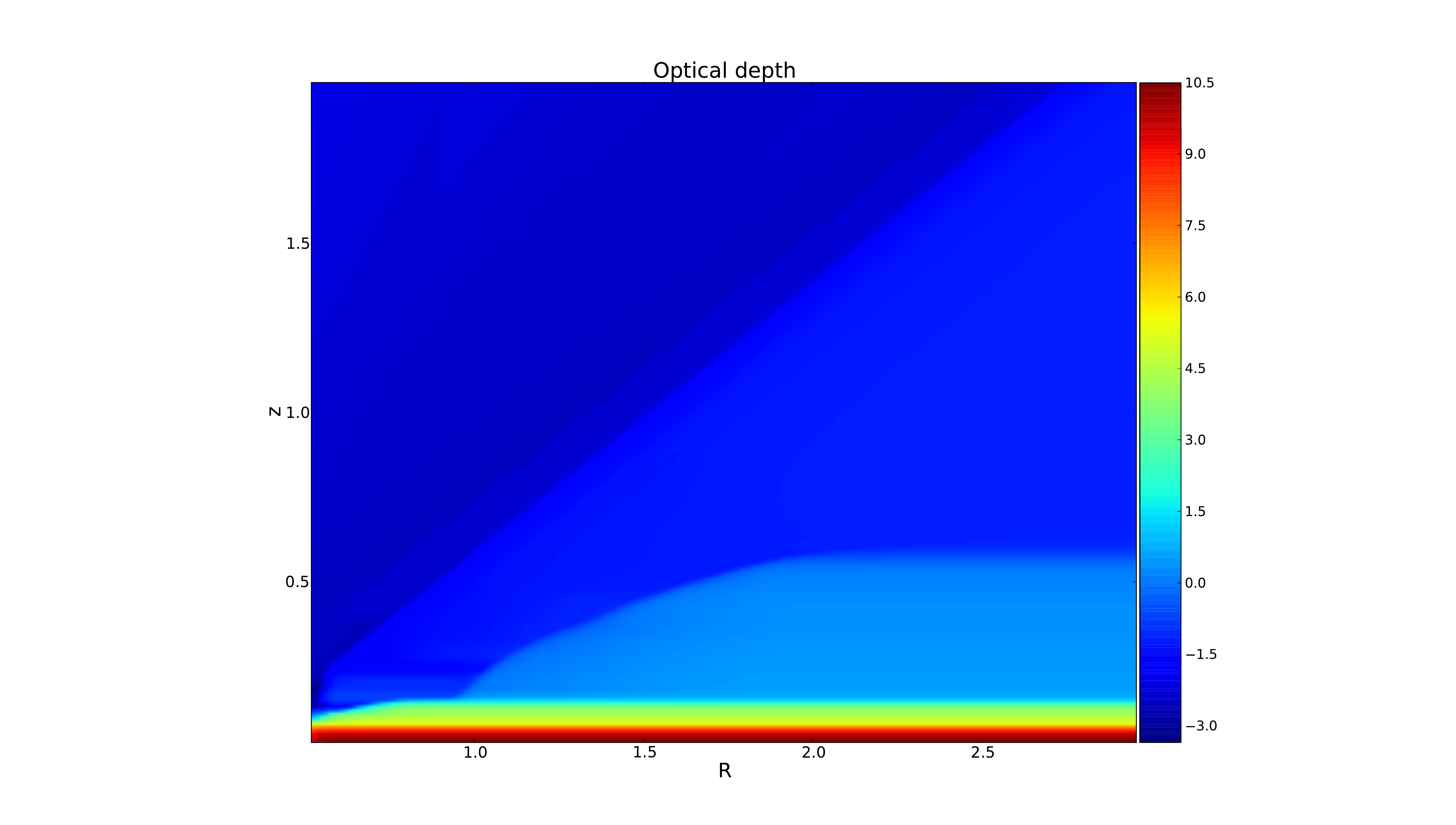}
\caption{
Color plot of the radial Thomson optical depth, $\log\tau$ from a BH towards a given point.
Axes: horizontal: $z$: distance from equatorial plane in parsecs; $R$: distance from the BH in parsecs;
}\label{figtau}
\end{figure}

\section{Mass-loss rate}
{
\mybf
The opacity of dusty plasma in the infrared domain is 10-20 times that of electron scattering
\citep{Semenov03}. Such huge opacity makes possible for the AGN to have a strong wind, radiating at a fraction of Eddington luminosity (defined with respect to Thomson scattering (see Paper I for an estimate of an outflow onset)  

Mass-loss rate from the IR driven portion of the wind can be estimated combining the results from a theory of stellar winds and some of the approximations adopted in Paper I.
From the stellar wind theory the mass-loss rate from a spherically symmetric wind can be found from the following approximate relation
\citep{LamersCassinelli99}:

\begin{equation}
{\dot M}/(4 \pi) \simeq  L_{\rm IR}/c \,\left(\Gamma_{\rm IR}-1)/(4 \pi  \, v^{\infty} \Gamma_{\rm IR})\right)\tau_{\rm w}
\mbox{,}
\end{equation}
where $\Gamma_{\rm IR}=L_{\rm X}/L_{\rm edd, IR}$, $v^{\infty}$ is the wind terminal velocity, and
$\tau_{\rm w}$ is the wind optical depth in IR. 
 
For simplicity, as in Paper I in this section we assume that all incident UV and X-ray radiation is reprocessed within a narrow conversion layer which effective temperature, $T_{\rm eff}$ can be found from the following approximate relation:
$\alpha\, \Gamma \eta_{\rm X} \, F_{\rm edd} = \sigma\, T_{\rm eft}^{4}$ where $\Gamma=L/L_{\rm edd}$, $\eta_{\rm X}=0.5$ is the fraction of X-ray radiation from the total radiation ,
and
the fraction $\alpha\simeq 0.5$ of the incident flux is re-emitted outwardly. 
Adopting the above, one obtains
$\Gamma_{\rm IR} = \alpha\, \Gamma \eta_{\rm X} \kappa_{\rm d}/\kappa_{\rm e}$ which
is approximately 
1.25 for $M_{\rm BH}=10^{7}\,\text{M}_\odot$,  $L=0.5 L_{\rm ddd}$ and $\kappa_{\rm d}=10 \kappa_{\rm e}$. The value of $v^{\infty}$ is  estimated
while neglecting $P_{\rm g}$ in favor of the radiation pressure: 
$
\displaystyle v^{\infty}=\sqrt{ 2 GM/r_{0} \,  (\Gamma_{\rm IR} -1)  } \simeq 207 \,
{\rm km\,s^{-1}  } 
$
adopting the above set of parameters plus the wind launching radius $R_{0}= 0.5 \, \text{pc}$,
and assuming that the wind occupies a wedge-like (in a quarter of the domain) space of an opening 
angle of $75^{\circ}$ we arrive to the estimate:
$ {\dot M} \simeq 3 \, \tau _w\,M_{\odot}\,\text{yr}^{-1}$. Monte-Carlo simulations show that the factor $\tau_{\rm w}$ can reach 1-5 for an AGN radiating at a fraction of $L_{\rm edd}$ \citep{Roth2012}, meaning the potential for the wind tori to reach 
mass loss rate of $10\,M_{\odot}\,\text{yr}^{-1}$; We do not see such mass-loss rates in our simulations, which may indicate that instead of being blown away as it would certainly do in a spherically-symmetric case, the gas is effectively ablated towards the equatorial plane by the pressure of the hot gas component. As the latter is itself the result of interaction with radiation, one can say that radiation effectively clears its way through the opening of the torus.

}

\section{Discussion}
The following average properties of our wind solutions can be summarized:

\begin{enumerate}
\item
The total mass-loss rate in the IR driven and warm absorber flows is between
$\langle {\dot M} \rangle  \simeq 0.1 M_{\odot}\, {\rm yr}^{-1}$ and $\langle {\dot M} \rangle  \simeq 1.5 M_{\odot}\, {\rm yr}^{-1}$ depending on the time of the evolution of the wind. 
This includes also the failed wind. The amount of gas which has enough energy to escape to infinity is in a good accord with the energetics derived from accretion.
Notice that approximately 
$1.6\times 10^{-1}-1.3\times 10^{-2}\,M_{\odot} \, {\rm yr}^{-1} $ 
are required to be accreted through
a geometrically thin accretion disk to support our models.
Also there is no problem of depletion of the torus since we have an infinite reservoir of matter in a razor thin equatorial disk.
\item
The IR driven-wind has the velocity in the range of of a few$\times 10-200\,{\rm km\,s^{-1}}$, and the fast
hot wind has velocity in the range of $400-800\,{\rm km\,s^{-1}}$.
\item
The transmission properties are similar between different models. The characteristic angle at which the wind becomes opaque to Thomson scattering is about $72^{\circ}-75^{\circ}$, and is determined by the balance between the hot, photo-ionized wind and the cold wind supported by the IR pressure on dust. 
It is interesting that the 
geometry of the torus is close to one obtained in \cite{Dorodnitsyn08b} at late times when the pressure of the hot, evaporative component becomes important.
\item
Both hot and cold components of the wind are non-stationary. 
\end{enumerate}

{\mybf 
Models such as
\cite{KoniglKartje94,ElitzurShlosman06,Emmering92,Everett05, Keating12, Bottorff2000} 
are developed on the premise of 1) the existence of a strong global magnetic field, and 2) uses the prescription of self-similarity. To a certain extent 
this limits the value of the direct comparison of our results with theirs.
Indirect comparison would include:
performing the detailed analysis of the spectral properties of the IR-driven ''wind torus'';
performing our simulations for the parameter range attributed to a well studied source (such as NGC5548, in case
of \cite{Bottorff2000}). Those are the next things to do in order to assess the validity of the current model and both of these goals is a natural continuation of the line of present studies.
}
We will elaborate on observational properties and the implications of the time-dependent behavior of our solutions in the following paper of this series.

\section{Conclusions}

Our studies of AGN infrared-driven winds have proceeded incrementally.
In Paper I we considered a toy model which roughly approximated an infrared driven obscuring wind in AGN. 
An important prediction which resulted was that if the temperature of the dusty plasma in the rotating torus
exceeds some characteristic value, the IR pressure on dust will inevitably overwhelm gravity, creating an outflow. However, this effective temperature, which approximately reads as
$ T_{\rm eff} = \left(\frac{GM\rho}{a r}\right)^{1/4} \simeq 312\,(\frac{n_{5} M_{7}}{r_{\rm pc}})^{1/4}-
987\,(\frac{n_{7}M_{7}}{r_{\rm pc}})^{1/4} \,{\rm K}\mbox{,} $
was derived analytically from a simplified model.


In Paper I we combined calculations of stationary 1D motion in the z-direction with 2D calculations of the radiation field. The latter was calculated adopting 2D flux-limited diffusion approximation.
In Paper II we relaxed the assumption of 1D stationary motion and adopted a 2.5D time-dependent picture. One important ''toy-model'' simplification still remained though: there was no X-ray radiation explicitly taken into account. Instead, we calculated the effective temperature of the gas at the boundary of the computational domain. In the present work we relaxed this assumption and included X-rays into the global picture explicitly. 

Despite these advances,  our work still does not consider the following potentially important physical processes:

The assumption of a razor thin disk is idealized.  An accretion disk at parsec scales is likely self gravitating, possibly affected by 
gravitothermal instability 
\citep[i.e.,][]{Gammie01}.  
This may provide a local source of
energy and maintain a Toomre parameter 
$\Omega_{\rm T}\sim 1$ \cite{Toomre64}. In a marginally gravitationally unstable thin disk, the non-axisymmetric {\mybf perturbations lead to angular momentum transport which can be described by an effective viscosity 
\citep[i.e.,][]{Paczynski78, Shlosman89, Shlosman90, Rafikov09}. }
Self-consistent models of geometrically thick disks in AGN at parsec scales call for the effects of self-gravity {\it and} radiation physics in 3D. This difficult endeavor can be a subject of future research.
Another complication is that the problem of parsec scale winds may be intrinsically connected 
with the problem of accretion at parsec scale. Relaxing the assumption of an equatorial thin disk 
and treating accretion adopting an effective $\alpha$ prescription
is a possible next step in the development of the current model. 

The assumption of a razor thin disk as a source of matter underpins the current study. Certainly such a distribution of matter is in many respects most unfavorable for the formation of radiation driven flow, owing to the geometry of X-ray illumination. Thus, we expect that if the wind does form for certain $\Gamma$ in the thin disk case, then we may expect that in less idealized circumstances it will also form. To relax the assumption of a razor thin disk one would need to abandon the assumption of equatorial symmetry and to include accretion disk into self-consistent consideration.

Starting from the Paper I we see that much of the massive wind doesn't leave the potential well of the BH. This conclusion is confirmed through Paper II, and in the present work as well.
The precise fate of this gas is beyond the scope of these works, but the results suggest the following:
if the accreted gas is rejected by the BH at least some of it will return for additional attempts. 

We conclude with a mix of results and ideas driven by our current studies:
\begin{enumerate}
\item
The distribution of radiation force inside the dusty component of the flow depends on the shape of the photosphere where X-rays are converted into IR and can only be calculated in multi-dimensional simulations. This work includes both X-rays and IR into 2.5D time-dependent, hydrodynamic simulations . 
\item
We do not observe a rotationally supported quasi-static torus in our simulations, although our simulated winds are likely to resemble some of the observed properties of such tori.
\item
The flow in the current simulations is more complex and time-dependent than that of our previous studies where we did not include X-rays explicitly. 
\item

If geometrically thick dusty obscuration develops then
a hot photo-ionized flow with velocities of $100-700\,{\rm km\, s^{-1}}$ accompanies it.  
X-ray warm absorbers are the evaporative flow originating both from the  disk 
and also evaporated from such cold dusty component. 

\item
We speculate that
large scale motions with $v_{z}=\pm (\text{a few})\times 100\,{\rm km\, s^{-1}}$ originating from a spatially varying wind such we have calculated may be seen in maser emission in nearby bright type II AGNs.
\end{enumerate}

This research was supported by an appointment at the NASA
Goddard Space Flight Center, administered by CRESST/UMD
through a contract with NASA, and by grants from the NASA
Astrophysics Theory Program 10-ATP10-0171.
G.B-K. acknowledges the support of from the Russian Foundation for Basic Research
(RFBR grant 11-02-00602).

\bibliography{BibList-tor7}{}

\section{Appendix A}
Frequency-integrated moments 
$E$, $\bf F$,  which appear in the above set of equations (\ref{eq11})-(\ref{eq14}) are obtained by
calculating angular moments from the frequency-integrated specific intensity, $I({\bf r}, \Omega, \nu, t)$:

\begin{eqnarray}\label{moments1}
E(\bf{r},t) &=& \frac{1}{c}\int_{0}^{\infty}d\nu  \oint d\Omega \, I({\bf r}, \Omega, \nu, t)\mbox{,}\\
{\bf F}(\bf{r},t) &=& \int_{0}^{\infty}d\nu\oint d\Omega \,\hat{\bf n} \,I({\bf r}, \Omega, \nu, t)\mbox{.}
\end{eqnarray}
The frequency-independent radiation pressure tensor
$\bf P$ is found from

\begin{equation}\label{moments2}
{\bf P}({\bf r},t) = \frac{1}{c}\int_{0}^{\infty}d\nu\oint d\Omega \,\hat{\bf n} \hat{\bf n} \,I({\bf r}, \Omega, \nu, t)\mbox{.}
\end{equation}

\section{Appendix B}
Dust directly reprocess X-rays to IR and we approximately take this into account. 
The amount of energy absorbed by dust in a volume $d^{3}x$:
\begin{equation}
\frac{1}{d^{3}x} \frac{dE^{\nu}_{\rm x}}{dt} = \alpha\, n_{\rm d}\ \int \sigma_{\rm x} I_{\nu\rm x} \, d\Omega\, d\nu 
\simeq \alpha\, n_{d } \sigma_{\rm x} c\, u_{x}\mbox{,}
\end{equation}
where $u_{\rm x}$ is the energy density of X-rays, $\sigma_{\rm x}=\chi_{\rm x}/n_{\rm d}$, and
the number density of dust grains reads: $n_{d} = f_{\rm d}\rho/m_{\rm d}$, where $m_{\rm d}$ is a mass of a grain, and  $f_{\rm d}$ is a dust-to-gas mass ratio. We make further simplifying assumption that the dust grain is being instantaneously heated to the effective temperature $T_{\rm d,x}$.

The contribution from the dust to the IR energy density (or, equivalently, to the temperature, $T_{\rm r}$ is calculated from the following equation:

\begin{equation}
\frac{d T_{\rm r}}{dt}=n_{\rm d}\,c\, \sigma_{d}\left({ T_{\rm d,x}} - { T_{\rm r}}\right)\mbox{.}
\end{equation}\label{dust_x_update}
Assuming, that $T_{\rm d,x}$ is constant over the time interval $[t, t+dt]$, the update for radiation is 
${T_{\rm d,x} }(t+dt) = {T_{\rm d,x}} (1-e^{- dt/t_{d}} )+ -e^{- dt/t_{d}} {T_{\rm d,x} }(t)$. Notice, that $E=aT_{\rm r}^{4}$.
The equation (\ref{dust_x_update}) is used to in operator-split fashion to update $E$ prior to all
other terms in equation (\ref{eq14}) are updated.

\end{document}